\documentclass[%
 apl,
 amsmath,amssymb,%
preprint,%
superscriptaddress,
]{revtex4-1}
\usepackage{array}
\usepackage{graphicx}% Include figure files
\usepackage{dcolumn}% Align table columns on decimal point
\usepackage{bm}% bold math
\usepackage{cleveref}
\usepackage{siunitx}
\usepackage{xcolor}

\newcommand{\BSMFO}{Ba$_{0.5}$Sr$_{1.5}$Mg$_{2}$Fe$_{12}$O$_{22}$}

\newcommand{\approptoinn}[2]{\mathrel{\vcenter{
  \offinterlineskip\halign{\hfil$##$\cr
    #1\propto\cr\noalign{\kern2pt}#1\sim\cr\noalign{\kern-2pt}}}}}

\newcommand{\appropto}{\mathpalette\approptoinn\relax}

\renewcommand{\vec}[1]{\mathbf{#1}}

\usepackage{lineno}

\begin{document}
%\linenumbers
%\preprint{}

\title[Title]{Magnetoelectric domains and their switching mechanism in a Y-type hexaferrite}

\author{F. P. Chmiel}
\affiliation{Clarendon Laboratory, Department of Physics, University of Oxford, Parks Road, Oxford OX1 3PU, United Kingdom}
\author{D. Prabhakaran}
\affiliation{Clarendon Laboratory, Department of Physics, University of Oxford, Parks Road, Oxford OX1 3PU, United Kingdom}
\author{P. Steadman}
\affiliation{Diamond Light Source, Harwell Science and Innovation Campus, Didcot OX11 0DE, United Kingdom}
\author{J. Chen}
\affiliation{Clarendon Laboratory, Department of Physics, University of Oxford, Parks Road, Oxford OX1 3PU, United Kingdom}
\author{R. Fan}
\affiliation{Diamond Light Source, Harwell Science and Innovation Campus, Didcot OX11 0DE, United Kingdom}
\author{R. D. Johnson}
\affiliation{Clarendon Laboratory, Department of Physics, University of Oxford, Parks Road, Oxford OX1 3PU, United Kingdom}
\author{P. G. Radaelli}
\affiliation{Clarendon Laboratory, Department of Physics, University of Oxford, Parks Road, Oxford OX1 3PU, United Kingdom}

\date{\today}

\begin{abstract}

By employing resonant X-ray microdiffraction, we image the magnetisation and magnetic polarity domains of the Y-type hexaferrite \BSMFO{}. We show that the magnetic polarity domain structure can be controlled by both magnetic and electric fields, and that full inversion of these domains can be achieved simply by reversal of an applied magnetic field in the absence of an electric field bias. Furthermore, we demonstrate that the diffraction intensity measured in different X-ray polarisation channels cannot be reproduced by the accepted model for the polar magnetic structure, known as the 2-fan transverse conical (TC) model.  We propose a modification to this model, which achieves good quantitative agreement with all of our data.   We show that the deviations from the TC model are large, and may be the result of an internal magnetic chirality, most likely inherited from the parent helical (non-polar) phase.   
\end{abstract}

\pacs{75.85.+t,}% PACS, the Physics and Astronomy
                             % Classification Scheme.
\keywords{Suggested keywords}%Use showkeys class option if keyword
                              %display desired
\maketitle

\section{Introduction}

The desire to reduce the thermal waste produced by modern IT components has driven an interest in magnetoelectric and multiferroic materials, which enable the control of magnetic order by electric fields \cite{eerenstein_2006}. In the case of `classic' magnetoelectric materials such as Cr$_2$O$_3$, the antiferromagnetic ordering pattern can be set by the simultaneous application of magnetic and electric fields, but the electrically-induced magnetisation is very small even for a single domain \cite{iyama_2013}.  By contrast, a key criterion for a range of potential applications is a very large magnetoelectric effect, coupled with the ability to control magnetoelectric domain patterns \cite{leo_2018}.  Of all the known magnetoelectric materials, the Y-type hexaferrites are among the closest to fulfil this criterion and have therefore attracted significant attention \cite{kimura_2012,zhai_2017,kocsis_2019}.

The Y-type hexaferrites exhibit complex magnetic phase diagrams as a function of temperature and magnetic field \cite{kimura_2012}.  In a specific range of fields and temperatures, a sizeable electrical polarisation is induced by the spin-current mechanism, coupled with a magnetoelectric coefficient that is many orders of magnitude larger than for Cr$_2$O$_3$ \cite{ishiwata_2008}. Notably, \BSMFO{} (BSMFO) possesses the largest known magnetoelectric coefficient of any material: it's sizeable electric polarisation can be switched, without any loss in magnitude, by small magnetic fields \cite{zhai_2017}. Magnetoelectric effects have also been demonstrated at room temperature in some Y-type hexaferrites \cite{hirose_2014}.  In spite of these desirable properties and after more than a decade of intense research, the precise nature of the magnetoelectric domain switching mechanism, as well as the microscopic details of the polar magnetic structures, are yet to be determined. 

In this article, we present the results of a Resonant X-ray Diffraction (RXD) experiment designed to study the magnetoelectric domain switching in BSMFO.  Historically, resonant X-ray microdiffraction (spatially resolved diffraction) has been employed to determine the real-space distribution of chirality domains, such as those of the zero-field, proper-screw phases of the Y-type hexaferrites, which are not magnetoelectric \cite{hiraoka_2011,ueda_2016,hiraoka_2015}.   Here, we demonstrate that this technique is also sensitive to the magnetic polarity of the field-induced polar phases, even in the absence of net chirality. Using this technique, we spatially resolve the magnetoelectric domain configuration of BSMFO for the first time.  We show that a reversal of the applied magnetic field results in reversal of the magnetic polarity, which is directly coupled to the electrical polarisation.  We also show that the domain boundaries in the system are toroidal in nature, are  stable upon magnetic field reversal, and can only be removed by simultaneous application of electric and magnetic fields. Polarisation analysis of the scattered X-rays with linear incident polarisation, performed dynamically during the magnetoelectric switching process, shows that the magnetoelectric switching occurs by collective rotation of the magnetic structure in spin space around the high-symmetry axis.  Moreover, the presence of an asymmetry between opposite directions of the magnetic field in the $\sigma \pi'$ and  $\pi \sigma'$ channels incontrovertibly demonstrates that the magnetoelectric phase cannot be a simple `transverse conical' structure, as previously proposed \cite{zhai_2017} and that the deviation from this structure required to explain the data quantitatively must be large.  We construct an explicit modification of the transverse conical structure, which fits all of our data well.  Based on this model, we speculate that the magnetoelectric phase could inherit internal chirality and a small ferrimagnetic $z$-axis component from the `parent'  non-polar helical/longitudinal conical phase.

\subsection{The magnetic structure of \BSMFO{}}

\begin{figure}
\centering
\includegraphics{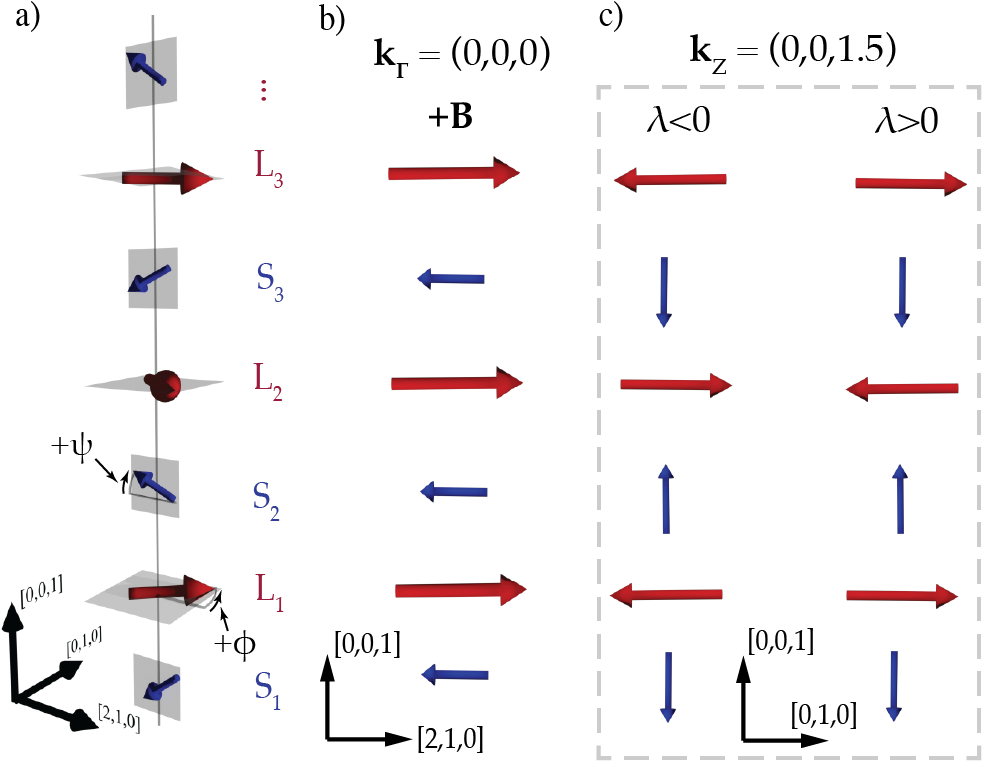}
\caption{a) 2-fan Transverse Conical phase (TC) magnetic structure within the block model. Red and blue arrows represent L and S block moments respectively, $\psi$ and $\phi$ indicate the respective fan angles of the blocks.  The applied magnetic field is parallel to the mean direction of the L blocks. b-c) Orthographic projection of the magnetic structure into two orthogonal planes, showing the TC structure can be considered as a linear combination of a ferrimagnetic component (b) and a commensurate cycloidal component (c), with respective propagation vectors of $\vec{k}_{\Gamma}  = (0,0,0)$ and $\vec{k}_{Z} = (0,0,1.5)$. Two possible magnetic polarity domains, arising because of the cycloidal component of the structure,  are displayed in c.}
\label{fig:canted_2fan}
\end{figure}

\BSMFO{} crystallises in space group $R\bar{3}m$ with lattice parameters $a=b\approx 5.8$  \si{\angstrom} and $c\approx 43$ \si{\angstrom}, having 6 symmetry-inequivalent magnetic Fe sites in the primitive unit cell.  Most of BSMFO's magnetic phases are complex and non-collinear, and an unconstrained magnetic structural solution of all but the simplest structures has thus far proven elusive even for single-crystal neutron diffraction \cite{momozawa_1993, lee_2012}.  Therefore, magnetic structures are almost universally described  by an approximate model, known as the \textit{block model} \cite{kimura_2012}. In this approximation, the unit cell is divided into two magnetic blocks, L (large moment) and S (small moment), which are sequentially stacked along $c$. Within each magnetic block, Fe moments are assumed to be collinear and ferrimagnetic, by virtue of strong nearest-neighbour superexchange \cite{utsumi_2007}.  The magnetic structure is then described by a set of `superspins', located at the centre of each block, each accounting for the net moment of the block. The net magnetic moments between neighbouring blocks can be collinear, yielding a globally ferrimagnetic phase, or non-collinear, giving rise to a plethora of helical and conical phases \cite{kimura_2012}.  In BSMFO, the net moments of the L and S blocks are 3$m_{Fe}$ and 1$m_{Fe}$ respectively \footnote{This assumes a random distribution of the Mg ions over the Fe tetrahedral and octahedral sites.}, where $m_{Fe}$ is the average moment magnitude of a single Fe site. In the first part of this paper, we will use the block model description of the magnetic structures (see \cref{fig:canted_2fan}, where L and S block moments are represented by red and blue arrows respectively).  Later, we will demonstrate that the `standard' block model fails to describe some of our RXD data, and discuss how the model needs to be modified on account of this.

At high temperatures, BSMFO is ferrimagnetically ordered, as discussed above, with anti-parallel magnetisations between the L and S blocks. At $\sim$ 390 K, the system undergoes a transition to an incommensurate \emph{planar} helical (proper screw) structure, in which neighbouring L and S superspins are aligned approximately antiparallel to one another, and coherently rotate in the $ab$ plane on traversing the $c$ axis \cite{chmiel_2018}.  Upon further cooling, the blocks are reported to cant out of the basal plane  to form what is known as the longitudinal conical (LC) structure, which was previously characterised in the Sr-free variant, Ba$_2$Mg$_2$Fe$_{12}$O$_{22}$ \cite{ishiwata_2010}. Application of small ($\sim$ 100 mT) in-plane magnetic fields in either of these non-collinear phases induces a series of `fan' phases \cite{momozawa_1993}, in which the blocks oscillate in the $ab$ plane around the direction of the net magnetisation.  In BSMFO and other hexaferrites, this in-plane fan oscillation is believed to be coupled with a modulation of the out-of-plane $c$ component, forming the so-called 2-fan / transverse conical (TC) structure \cite{ishiwata_2008, kimura_2012}. The TC phase is electrically polar, with its cycloidal component  believed to be responsible for the induced electrical polarisation through the spin current mechanism (see \cref{sec:polarisation_measurements}).  The TC structure of relevance to this paper is stabilised below 100\,K by application of a 150\,mT magnetic field applied parallel to [2,1,0] ($a^*$). It is described by two propagation vectors, $\vec{k}_{\Gamma} = (0,0,0)$ and $\vec{k}_{Z} = (0,0,1.5)$  (see Appendix \ref{Appendix: TC model} or \cite{kimura_2012} for details). % as shown in \cref{fig:canted_2fan} a

\subsection{Magnetic polarity and electrical polarisation}
\label{sec:polarisation_measurements}
According to the spin-current mechanism, the electrical polarisation of a non-collinear phase is given by \cite{katsura_2005},

\begin{equation}
 \vec{P}\propto \hat{e}_{i,j}\times(\vec{S}_i \times \vec{S}_j),
 \label{eq:polarisation}
\end{equation}

where $S_{i,j}$ are neighbouring spin blocks and $\hat{e}_{i,j}$ the unit vector joining them. In the TC phase the electric polarisation is induced by the cycloidal  component of the magnetic structure ($\vec{k}_{Z} \parallel \vec{c}$). This results in an electrical polarisation $\vec{P}$ in the plane of rotation of the spins and perpendicular to both $\vec{k}_{Z} $ and  $\vec{B}$ \cite{zhai_2017}, and hence to the magnetisation $\vec{M}$. For the TC phase we can write the amplitude of $\vec{M}$ as 
\begin{equation}
m=m_L\cos \phi -m_S\cos \psi,
\end{equation}
where $m_i$ are the net moment magnitudes of the blocks and $\phi$ and $\psi$ the block fan angles (see \cref{fig:canted_2fan}). Since $\vec{P}$ and $\vec{M}$ are orthogonal, the TC phase also possesses a \emph{ferrotoroidal moment} $\vec{T}$, defined as:
\begin{equation}
\label{eq: ferrotoroidal_definition}
\vec{T}=\vec{P} \times \vec{M}.
\end{equation}

Here, the ferrotoroidal moment defines a specific type of magneto-electric domain, which will be discussed in detail later.

A peculiar feature of BSMFO is that the sign of the electric polarisation can be easily switched by reversing the sign of the magnetic field used to stabilise the TC phase \cite{ishiwata_2008, ishiwata_2008,zhai_2017}. From inspection of \cref{eq:polarisation}, one can see that reversal of the electric polarisation could be accomplished by a global rotation of the magnetic structure in spin space around the axis parallel to $\hat{e}_{i,j}$. In the analysis that follows, it is useful to introduce a magnetic order parameter $\bm{\lambda}$, known as the \emph{magnetic polarity}, which provides a direct coupling between the magnetism and the electrical polarisation of the form $\vec{P} = \chi\bm{\lambda}$, where $\chi$ is a coupling constant \cite{johnson_2013}. $\bm{\lambda}$ is quadratic in the magnetic moments, parallel to $\vec{P}$, and in the TC phase its amplitude, $\lambda$, can be written as,
\begin{equation}
\lambda = m_L m_S \sin \phi \sin \psi.
\label{eq:magnetic_polarity}
\end{equation}

Two possible magnetic polarity domains are displayed in \cref{fig:canted_2fan} c. We have directly imaged the magnetic polarity of BSMFO by RXD and, by virtue of the linear relation between the magnetic polarity and electric polarisation, also  established the electric polarisation domain configuration.

\section{Experimental Details}
\label{sec:experimental}
% Basic experimental introduction: Crystal used in study

\subsection{Crystal growth and characterisation}
\label{section: xtal growth char}

Single crystals of \BSMFO{} were grown by the flux method \cite{momozawa_2001}. Laboratory-based X-ray diffraction and magnetisation measurements were performed (presented in the supplementary information) to check the crystal quality used in our RXD experiment. % Magnetometry was performed by mounting a 0.38\,mg sample into a Quantum Design MPMS-3 aligned such the [2,1,0] direction was parallel to the applied magnetic field. The sample was zero-field cooled to the measurement temperature from 400\,K for each measurement.%with lattice parameters of $a = 5.857 \AA$ $c = 43.353 \AA$ in space group R$\bar{3}$m

To investigate the polar properties of the sample, magnetocurrent measurements were performed using a custom probe inserted into a Quantum Design PPMS. The sample was mechanically polished into a cuboid and silver paste was used to make electrical contacts on the $b$-faces, the [2,1,0] direction (orthogonal to the $b$-faces) was oriented parallel to the applied magnetic field. Samples were zero field cooled from room temperature to 10\,K and then an electric field bias (of magnitude 100\,Vmm$^{-1}$) was applied to the sample. The magnetic field was increased to $+0.5$\,T or $-0.5$\,T, at which point the biasing electric field was removed. The magnetic field was then swept whilst measuring  the displacement current, the integral of which yields the sample polarisation as a function of the applied magnetic field. 

\begin{figure}[!b]
\includegraphics{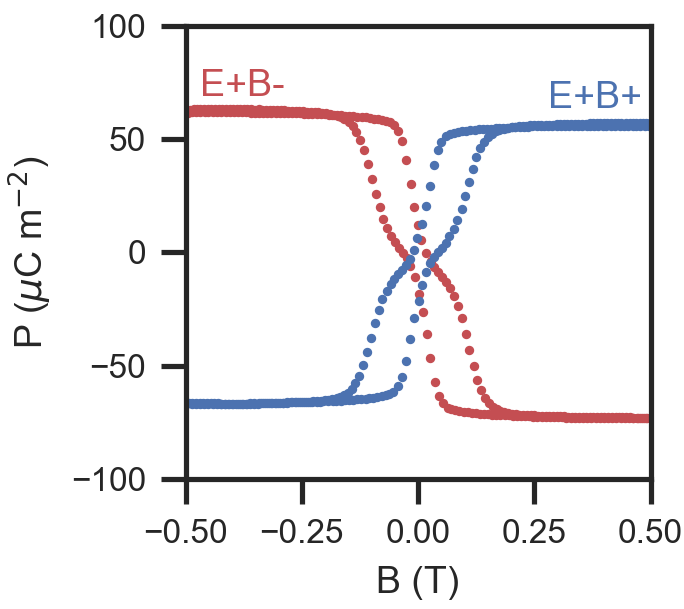}
\caption{Electric polarisation of BSMFO as a function of applied magnetic field, following different magnetoelectric poling procedures. Measurements (color) were performed in the absence of an applied electric field at 10 \,K. Blue circles indicate measurement following a magnetoelectric poling with a E$_+$B$_+$ poling field product (see main text) and red circles with a E$_+$B$_-$ field product. The magnitude of E$_+$ was 100 \,Vmm$^{-1}$.}
\label{fig:ME_measurements}
\end{figure}

By using this procedure, the electric polarisation and its switching behaviour were fully characterised at 10\,K (\cref{fig:ME_measurements}). Initially, the sample was prepared into a single polar domain  using a E$_+$B$_+$ `poling' cross fields. After the removal of the electric field, the electric polarisation (blue circles in \cref{fig:ME_measurements}) was measured as the magnetic field was swept from $0.5$\, T to $-0.5$\, T and back. The measured polarisation is displayed by the blue circles in \cref{fig:ME_measurements}. We observe an appreciable electric polarisation \footnote{We found this to be sample dependent and could be as large in magnitude as 200 $\mu Cm^{-2}$} which, importantly, is coupled to the direction of the applied magnetic field. Consistent with previous studies on BMFO \cite{ishiwata_2008,taniguchi_2008}, we found the reversal of the electric polarisation to be a lossless process: i.e., the polarisation direction can be switched without any loss in magnitude, even in the absence of an electric field bias.

Secondly, we investigated the effect that different magnetoelectric poling fields had on our samples. The solid red points in \cref{fig:ME_measurements} show the measured electric polarisation after poling the sample with E$_+$B$_-$ poling fields. In this situation, we see that the direction of the electric polarisation is anti-parallel to the E$_+$B$_+$ case, is of similar magnitude, and is coupled to the direction of the applied magnetic field in the same way. These observations confirm that our sample exhibits the magnetoelectric effects of interest.

\subsection{Resonant X-ray diffraction: experimental geometry}

\begin{figure}
\centering
\includegraphics{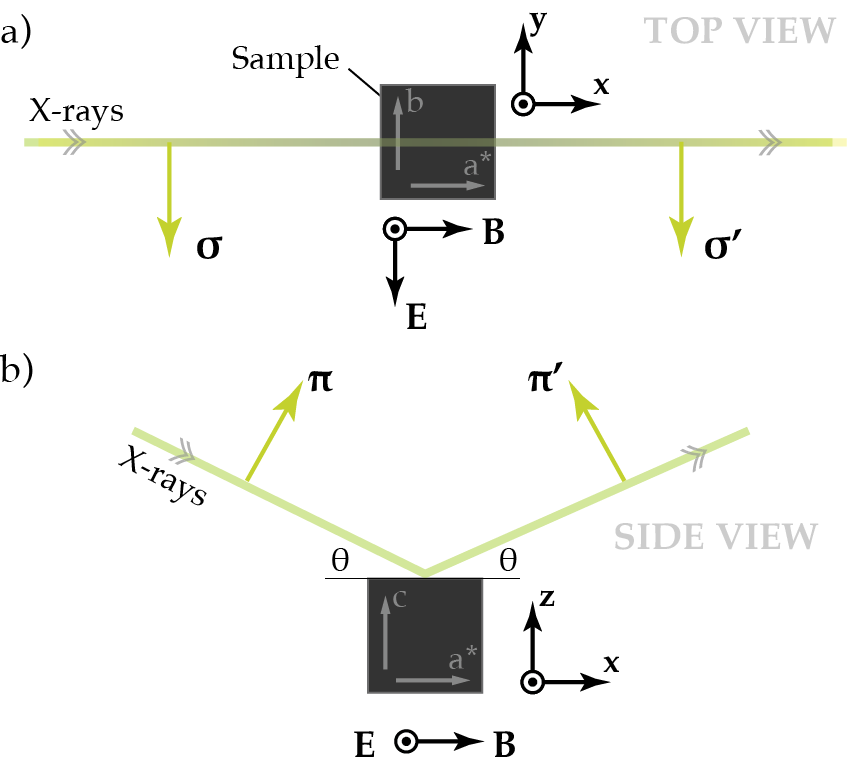}
\caption{Graphic demonstrating the geometry of the resonant X-ray diffraction experiment. a) Orthographic projection down the crystallographic $c$ axis. b) Projection parallel to the $b$ axis. The sample sits at the centre of two pole pieces of an electromagnet, which can apply magnetic fields parallel to the scattering plane ($\parallel x$). Electric fields are applied orthogonal to the scattering plane ($\parallel y$).  The basis ($\sigma-\pi$) of orthogonal linear polarisation states are shown by solid yellow arrows and the Cartesian coordinate system used in our calculations is displayed by the inset axes. The crystallographic (hexagonal) basis is shown by the light grey arrows overlaid on the sample graphic. } 
\label{fig:RXS_geometry}
\end{figure}

Resonant soft X-ray diffraction experiments were performed using the RASOR diffractometer at the I10 beamline located at the Diamond Light Source (Didcot, UK) \cite{beale_2010}. The experimental geometry is shown in fig. \ref{fig:RXS_geometry}, which includes the Cartesian basis used throughout this paper. The sample was cleaved in ambient conditions, yielding smooth faces normal to $c~(z)$ and mounted such that $c$ and $a^*~(x)$ were orientated in the scattering plane, with $c$ in the specular direction. An electromagnet and  a $^4$He flow cryostat were used to apply magnetic fields between $\pm$ 300 \si{\milli\tesla} parallel to $a^*$ at sample temperatures between 25 K and 300 K. The incident X-ray energy was set to 707.48 eV (the Fe $L_3$  absorption edge) for the duration of the experiment. At this energy, the X-ray attenuation length is approximately 0.5 \si{\micro\metre} \cite{hearmon_2013}. When an \textit{in situ}  electric field was required, silver paste was used to make contacts on opposite faces of the sample, such that electric fields could be applied mutually perpendicular to the applied magnetic field and the $c$ axis.

Diffraction measurements were performed along the $(0,0,l)$ zone axis. In our experimental geometry, the Ewald sphere only contains the $(0,0,3)$ Bragg peak, with contributions from Thomson (non-magnetic) scattering as well as magnetic scattering from the net magnetisation of the sample, the two magnetic satellites $(0,0,3 \pm k_{\Lambda})$ of the helical phase in zero field ($k_{\Lambda} \sim 0.6-0.8$), and the $(0,0,4.5)$ or $(0,0,1.5)$ purely magnetic peaks of the TC phase. The reciprocal space coordinates $(0,0,3 \pm 2k_{\Lambda})$, which correspond to higher-order scattering processes, were also accessible (see below). 

Two perpendicular, linear x-ray polarisations of the incident beam could be selected, denoted $\epsilon_\sigma$ and $\epsilon_\pi$ (fig. \ref{fig:RXS_geometry}), as well as circular positive $(c^+)$ and circular left $(c^-)$ polarisations. The linear polarisation of the diffracted beam could be determined using a multilayer analyser whose $d$ spacing was selected to achieve scattering at Brewster's angle when using an incident X-ray energy of 707.48 eV.

X-ray microdiffraction was achieved using a 50\,\si{\micro\metre} diameter aperture placed 265 mm upstream of the sample.

\section{Resonant X-ray scattering cross sections}

\begin{table*}
\centering
\caption{\label{table: magnetic SF} Magnetic interaction vectors for the phases and Bragg reflections considered in this paper. Helix/LC: Helical/longitudinal conical phase.  TC: 2-fan transverse conical phase. m-TC: modified  2-fan transverse conical phase.  The Bragg angles of the commensurate 2-fan phases at low temperatures and at the photon energy employed for the experiment ($\hbar \omega =$707.48 eV, with a wavelength of 17.5250 \AA) are also indicated. For the helical/LC phase, the superscripts refer to the direction of the magnetic moment parallel or perpendicular to the $xy$-plane (N.B. the perpendicular component is zero in the planar helical phase).}
\label{tab:magnetic interaction vectors}
    \begin{ruledtabular}
    \begin{tabular}{c | c  c  c }
    &\multicolumn{3}{c}{Reflection: $(0,0,3-k)$/$(0,0,1.5)$, $\theta_{(0,0,1.5)}=17.67^{\circ}$}  \\[0.5ex]
    \hline\\[-2ex]
    Phase&$M_x$&$M_y$&$M_z$\\
     \hline\\[-2ex]
    Helix/LC&$ -\frac{1}{2} (m^{\parallel}_L+m^{\parallel}_S)$&$\frac{1}{2} i (m^{\parallel}_L-m^{\parallel}_S)$&$0$\\
     \hline\\[-2ex]
    TC&$0$&$i m_L \sin \phi $&$m_S \sin \psi $\\
    \hline\\[-2ex]
    m-TC&$m_S \sin \psi  \sin \psi _1$&$ i m_L \sin \phi  \cos \phi _1$&$m_S \sin \psi \cos \psi _1$\\
      \hline\\[-2ex]
      \hline\\[-2ex]
     &\multicolumn{3}{c}{Reflection: $(0,0,3+k)$/$(0,0,4.5)$, $\theta_{(0,0,4.5)}=66.30^{\circ}$}  \\[0.5ex] 
     \hline\\[-2ex]
        Phase &$M_x$&$M_y$&$M_z$\\
     \hline\\[-2ex]
    Helix/LC&$-\frac{1}{2} (m^{\parallel}_L+m^{\parallel}_S) $&$ -\frac{1}{2} i (m^{\parallel}_L-m^{\parallel}_S) $&$0$\\
     \hline\\[-2ex]
    TC&$0$&$-i m_L \sin \phi $&$m_S \sin \psi$\\
    \hline\\[-2ex]
    m-TC&$m_S \sin \psi  \sin \psi _1$&$ -i m_L\sin \phi  \cos \phi _1$&$m_S \sin \psi  \cos \psi _1$\\
      \hline\\[-2ex]
      \hline\\[-2ex]
     &\multicolumn{3}{c}{Reflection: $(0,0,3)$, $\theta_{0,0,3}=37.62^{\circ}$}  \\[0.5ex] 
     \hline\\[-2ex]
         Phase&$M_x$&$M_y$&$M_z$\\
     \hline\\[-2ex]
    Helix/LC&$0$&$0$&$ m^{\perp}_S-m^{\perp}_L$\\
     \hline\\[-2ex]
    TC&$m_L \cos \phi -m_S \cos \psi $&$0$&$0$\\
    \hline\\[-2ex]
    m-TC&$m_L \cos \phi  \cos \phi _1-m_S \cos \psi  \cos
   \psi _1 $&$0$&$m_S \cos \psi \sin \psi _1 $\\
    \end{tabular}
    \end{ruledtabular}
\end{table*}

In this section, we summarise key equations for the magnetic scattering cross sections in the linear polarisation channels, and other important derived quantities such as circular dichroism. The equations are given in terms of the magnetic interaction vector, $M_\gamma$, used extensively in the analysis of magnetic neutron diffraction data, and $F^{(0)}$ and $F^{(1)}$, the resonant scattering factors that are in general complex numbers \cite{hill_1996}. We note that there exists an additional resonant scattering factor $F^{(2)}$, which for an incommensurate magnetic structure having propagation vector $\vec{k}$, gives rise to magnetic satellites at $Q=(003)\pm 2 \vec{k}$, with an intensity proportional to $|F^{(2)}|^2$.  For the helical phase of BSMFO, we measured an intensity ratio of $\le 5\%$ between $2 \vec{k}$ and $\vec{k}$ satellites, indicating that the $|F^{(2)}|$ is small compared to $|F^{(0)}|$ and is at most $\sim 20 \%$ of $|F^{(1)}|$ \cite{hill_1996}.  Nevertheless, for a commensurate structure, $F^{(2)}$ could provide a sizeable contribution to the magneto-structural interference at the $\Gamma$ point as well as $F^{(1)}$ -- $F^{(2)}$ interference, which we address in Section \ref{sec::beyond_block}. A complete summary of the theory of resonant magnetic scattering from which the following equations are derived is provided in Appendix \ref{Appendix: basic MS theory}, in which we only consider magnetic dipole contributions.

\begin{itemize}
\item The \emph{circular dichroism} is defined as

\begin{equation}
\delta  = \frac{I_+ - I_-}{I_+ + I_-}
\label{eq:circ_dichroism}
\end{equation}

where $I_+$ and $I_-$ are the total scattered intensities with incident $c^+$ and $c^-$ circular polarisation and without any polarisation analysis of the scattered beam.

\item The \emph{circular asymmetry}, i.e., the asymmetry of the circular dichroism upon reversal of the magnetic field ($B+ \rightarrow B-$, assuming that the $x$ and $y$ components are reversed), defined as

\begin{equation}
Asy_{\delta}  = \frac{\delta(B+)+\delta(B-)}{\delta(B+)-\delta(B-)}
\label{eq:circ_asymmetry}
\end{equation}

\item The \emph{linear asymmetries}, i.e., the asymmetry of the scattering in the linear channels upon reversal of the magnetic field, defined as:

\begin{equation}
Asy_{\epsilon \epsilon'}  = \frac{I_{\epsilon \epsilon'} (B+)-I_{\epsilon \epsilon'}(B-)}{I_{\epsilon \epsilon'} (B+)+I_{\epsilon \epsilon'}(B-)}
\label{eq:linear_asymmetry}
\end{equation}

As we shall see, the asymmetries are defined in such a way to be zero in the simplest versions of the magnetic structures, hence the opposite signs in equations \ref{eq:circ_asymmetry} and \ref{eq:linear_asymmetry}.

\end{itemize}

\subsection{Structural-magnetic interference: circular dichroism at the $\Gamma$ point}

For phases with a net magnetisation, one would expect to observe circular dichroism at the $\Gamma$ point (the (003) reflection in our case) since the Thomson and magnetic contributions interfere at this position \cite{pollmann_2000}.  For the simplest case in which the magnetic interaction vector is a real vector of amplitude $m$, directed along the $x$ axis (this is always the case for the phases considered here), we have:

\begin{equation} 
\label{eq: gamma_pt_dichro}
\delta^{(\Gamma)}=\frac{4 m\Re (F^{(0)} \overline{F^{(1)}})  \cos^3 \theta }{|F^{(0)}|^2 \left(\cos
   ^2(2 \theta )+1\right)+2 |F^{(1)}|^2 m^2 \cos ^2(\theta )}
\end{equation}

where the overline indicates complex conjugation and $\theta$ is the Bragg angle. $\Re()$ and $\Im()$ denote real and imaginary parts, respectively.  For this simple magnetic interaction vector, the dichroism changes sign upon reversal of the magnetisation, and consequently $Asy_{\delta}^{(\Gamma)}=0$.

\subsection{Circular dichroism for purely magnetic peaks}

The general form of the circular dichroism for a purely magnetic peak is

\begin{equation}
\label{eq: magnetic_dichro}
\delta=\frac{ \Im (M_y (\overline{M_x} \cos \theta + \overline{M_z} \sin \theta)) \sin (2 \theta)}{|M_x|^2 \cos ^2 \theta + |M_z|^2 \sin^2 \theta + |M_y|^2 \sin^2 (2 \theta)}
\end{equation}

A very important result can be obtained by inspecting eq. \ref{eq: magnetic_dichro}: the circular dichroism of a purely magnetic Bragg peak is always zero for centrosymmetric magnetic phases.  In these cases, the magnetic interaction vector is always a real vector multiplied by a global phase, and the numerator of \cref{eq: magnetic_dichro} is identically zero.  As we shall see, the circular dichroism is proportional to the magnetic chirality for the helical phase and to the magnetic polarity for the TC phase.

The circular dichroism averaged between the two directions of $B$ is:
\begin{eqnarray}
\label{eq: av_circ_dich}
\delta_{ave}&=&\frac{1}{2}\left(\delta_{B^+}-\delta_{B^-}\right)\nonumber\\
&=&\frac{\Im (M_y \overline{M_z}) \sin \theta \sin (2 \theta)}{|M_x|^2 \cos ^2 \theta + |M_z|^2 \sin^2 \theta + |M_y|^2 \sin^2 (2 \theta)}
\end{eqnarray}

The circular asymmetry is

\begin{equation}
\label{eq: magnetic_circ_asy}
Asy_{\delta}=\frac{\delta_{B^+}+\delta_{B^-}}{\delta_{B^+}-\delta_{B^-}}=\frac{\Im (M_y \overline{M_x}) \cot \theta}{\Im(M_y\overline{M_z})}
\end{equation}

\subsubsection{General formulas for the linear polarisation channels}

The scattering intensities in the linear channels are, to within a multiplicative factor,

\begin{eqnarray}
\label{eq: int_linear}
I_{\sigma \sigma'}&=&|F^{(0)}|^2\nonumber\\
I_{\sigma \pi'}&=&|F^{(1)}(M_x \cos \theta + M_z \sin \theta)|^2 \nonumber\\
I_{\pi\sigma'}&=&|F^{(1)}(M_x \cos \theta - M_z \sin \theta)|^2 \nonumber\\
I_{\pi\pi'}&=&|F^{(0)} \cos (2 \theta)-iF^{(1)}M_y \sin (2 \theta)|^2
\end{eqnarray}

where, for purely magnetic peaks, we can take $F^{(0)}=0$.

The linear asymmetries are:

\begin{eqnarray}
\label{eq: magnetic_linear_asy}
Asy_{\sigma \sigma}&=&0\nonumber\\
Asy_{\sigma \pi'}&=&\frac{\Re (M_x \overline{M_z}) \sin (2 \theta)}{|M_x|^2 \cos ^2 \theta +|M_z|^2 \sin ^2 \theta} \nonumber\\
Asy_{\pi\sigma'}&=&-\frac{\Re (M_x \overline{M_z}) \sin (2 \theta)}{|M_x|^2 \cos ^2 \theta +|M_z|^2 \sin ^2 \theta} \nonumber\\
Asy_{\pi\pi'}&=&\frac{\Im(F^{(1)}\overline{F^{(0)}} M_y) \sin(4 \theta)}{|F^{(0)}|^2 \cos ^2 (2 \theta) + |F^{(1)}|^ 2 |M_y|^2 \sin ^2 (2 \theta)}\nonumber\\
\end{eqnarray}

The last term is, once again, due to magnetic-structural interference; for purely magnetic peaks, we can set $F^{(0)}=0$, so that also $Asy_{\pi\pi'}=0$ .

%up to here

\subsection{Magnetic interaction vector calculations}

Using the block model approximation, the complex $M_x$, $M_y$ and $M_z$ components of the magnetic interaction vectors for different Bragg peaks can be easily calculated (this is performed in detail in Appendix \ref{Appendix: magnetic_models}).  The results are summarised in Table \ref{table: magnetic SF} for all the relevant reflections, together with the Bragg angles at which these reflections occur at the photon energy of 707.48 eV.  This information is sufficient to calculate all relevant intensities (to within a global scale factor), as well as the circular dichroism and asymmetries, using the formulas from the previous section.  

\section{Results and Discussion}

%\begin{figure}
%\centering
%\includegraphics{figures/characterisation.png}
%\caption{a) and b) Magnetic moment, measured at 10\,K and 50\,K respectively, as a function of applied magnetic field. Grey dashed line highlights the initial increasing field measurement, following a zero field cool to the measurement temperature from 300\,K. Blue solid lines are subsequent measurements. c) (0,0,l) scans measured in zero applied magnetic field (solid red line) and in 0.25\,T (solid blue line), in the helical and canted 2-fan phase respectively. For all panels, magnetic fields were applied parallel to [2,1,0].}
%\label{fig:characterisation}
%\end{figure}

\subsection{Helical phase}

\begin{figure}
\includegraphics{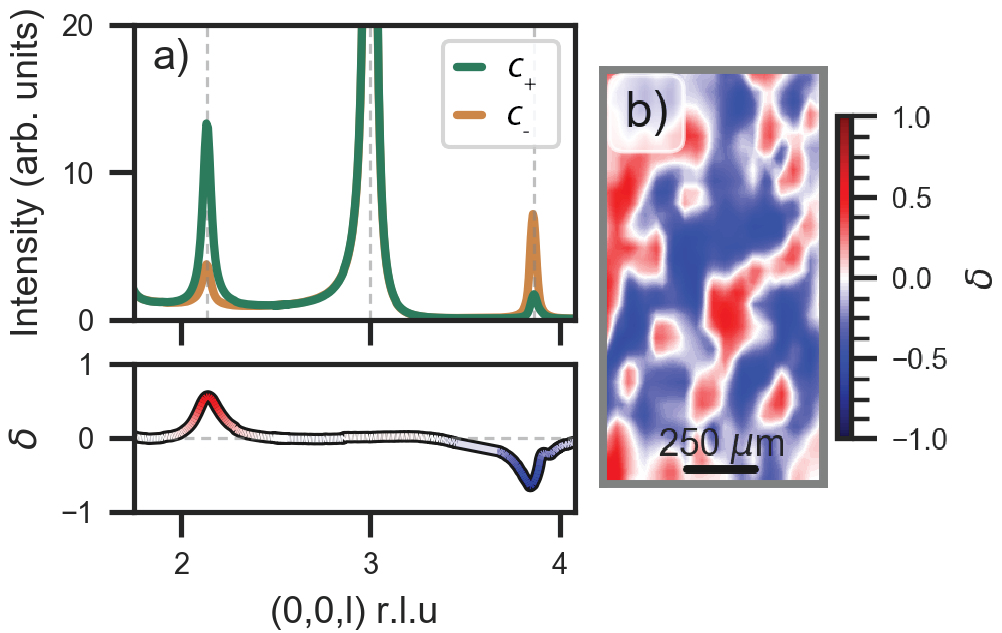}
\caption{Diffraction measurements of the helical phase. a) (0,0,l) scan measured with circular positive (solid green line) and circular negative (solid brown line) at 300 K in the absence of an applied magnetic field. The bottom panel displays the circular dichroism calculated from these measurements. Note the lack of appreciable dichroism for the (0,0,3) reflection and the large dichroism, of opposite sign, for (0,0,3$\pm$k). b) Spatially resolved circular dichorism (color) of the (0,0,3-k) reflection at 25 K in zero applied magnetic field.}
\label{fig:helical_domains}
\end{figure}

We begin by discussing our characterisation of the helical magnetic structure, stable in the absence of an applied field and for T $\le 390$ K. For this phase, the expectation from the previous section is that the (0,0,3) reflection should not be circular dichroic (there is no magnetic contribution to the scattering intensity at this position in this phase). Conversely, the two magnetic satellites, (0,0,3$\pm$k), should exhibit circular dichorism with magnitude equal to, 

\begin{equation}
\delta_{helix}=\frac{\pm2\gamma\sin \theta}{1 + 2 \sin ^2 \theta},
\end{equation} 

where $\pm$ denotes the measured satellite ($\pm$\textbf{k}) and $\gamma$ is the mean chirality of the probed region. Hence, the dichroism of the two satellites has opposite sign for domains of fixed chirality ($\gamma$).

\Cref{fig:helical_domains}a shows a (0,0,l) scan of the helical phase with circular incident X-rays, which indeed demonstrates opposite circular dichroism for the two incommensurate magnetic satellites. Furthermore, we spatially resolved the circular dichroism of the (0,0,3-k) reflection (\cref{fig:helical_domains} b) which exhibits clear regions of contrast. These two regions correspond to the two possible chirality domains of the helical state. The domain  morphology is similar to previous observations of chirality domains in other Y-type hexaferrites \cite{hiraoka_2011}. We also found that the measured helical domain configuration is invariant upon cooling to 25 K and to the application of small applied magnetic fields. It is reported that the helical phase transforms to a longitudinal conical state at low temperatures \cite{zhai_2017,ishiwata_2010}, which would imply that the (0,0,3) reflection should develop a small amount of dichroism.  However, no evidence of this transition was observed in our experiment for temperatures above 25 K.  In the following we refer to this phase as the helical phase, but remark that classification of this phase as a longitudinal state would not affect the presented results.

\subsection{Magnetoelectric TC phase at low temperatures}

Below approximately 100 K, the application of a 150 mT magnetic field in the basal plane of the helix transforms this phase to the TC state (see supplementary information for supporting magnetometry measurements). This transition was evident in our diffraction data by the disappearance of the two incommensurate magnetic satellites of the helical phase, and the emergence of two commensurate magnetic reflections at (0,0,1.5) and (0,0,4.5), both of which are characteristic of the TC phase \cite{zhai_2017}.

\subsubsection{Circular dichroism measurements}

\begin{figure}
\centering
\includegraphics{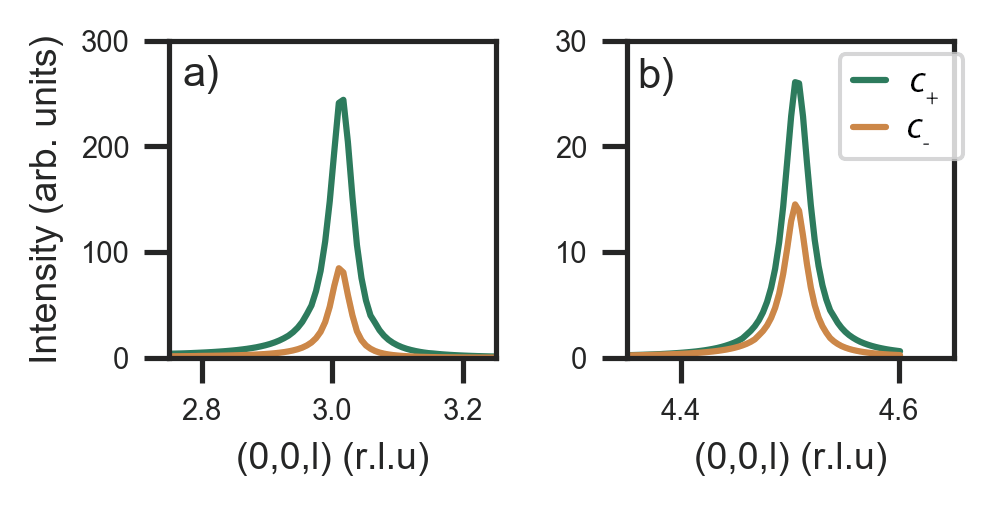}
\caption{Intensity of the (0,0,3) and (0,0,4.5) reflections (a and b respectively) measured  with circularly positive ($c_+$) and circularly negative ($c_-$) polarised incident X-rays. No polarisation analysis of the scattered beam was performed.  Measurements were performed at 40\,K and with a 25\,mT magnetic field applied parallel to [2,1,0].}
\label{fig:00l_circ_scans}
\end{figure}

The intensities of the (0,0,3) and (0,0,4.5) reflections were measured with circularly polarised incident X-rays of both polarities ($c^\pm$), as shown in \cref{fig:00l_circ_scans}. For these measurements, polarisation analysis of the scattered beam was not performed. Clearly, both the (0,0,3)  and (0,0,4.5) reflections exhibit a large difference in scattering intensity between $c^+$ and $c^-$ polarised incident X-rays and both, therefore, possess a large circular dichroism (\cref{eq:circ_dichroism}).
By inserting the appropriate magnetic interaction vectors and Bragg angles for the TC phase into equations \ref{eq: gamma_pt_dichro} and \ref{eq: magnetic_dichro}, we obtain the following expressions:

\begin{align}
\delta_{(003)}&=\frac{4m \Re (F^{(0)} \overline{F^{(1)}}) \cos^3 \theta}{|F^{(0)}|^2 (\cos^2 (2 \theta)+1)+2 |F^{(1)}|^2m^2 \cos^2 \theta}\nonumber\\
\delta_{(001.5)}&=-\frac{2 \lambda \cos \theta}{m_S^2 \sin^2 \psi+ 2m_L^2 \sin^2 \phi^2 \cos^2 \theta}\nonumber\\
\delta_{(004.5)}&=\frac{2  \lambda \cos \theta}{m_S^2 \sin^2 \psi+ 2m_L^2 \sin^2 \phi^2 \cos^2 \theta}.
\label{eq: dichroism_calculations_TC}
\end{align}

We conclude that, for this structural model, $\delta^{(003)}$ is to first order proportional to the amplitude of the net magnetisation, $m$, while $\delta^{(001.5)}$ and $\delta^{(004.5)}$ are proportional to the amplitude of the magnetic polarity $\lambda$ (\cref{eq:magnetic_polarity}). 

We emphasise that the circular dichroism of the two magnetic peaks naturally arises due to the non-centrosymmetric nature of the TC phase \cite{hiraoka_2011}, and there is therefore no need to invoke a hypothetical structural modulation with the same propagation vector \cite{ueda_2016}.  The absence of any scattering intensity at the (0,0,4.5) and (0,0,1.5) positions for scans performed with linear incident polarisation in the $\sigma \sigma'$ channel, which is only sensitive to the crystal structure, confirm that such a structural modulation is not present (see \cref{sec:lin_pol_analysis}). 

To investigate how the circular dichroisms vary over the extent of the sample we restricted the diameter of the incident X-ray beam to 50\,$\mu$m and performed a raster scan in steps of  50\,$\mu$m to  spatially resolve the circular dichroism of the (0,0,3) and (0,0,4.5) reflections.

\begin{figure}[!b]
\centering
\includegraphics{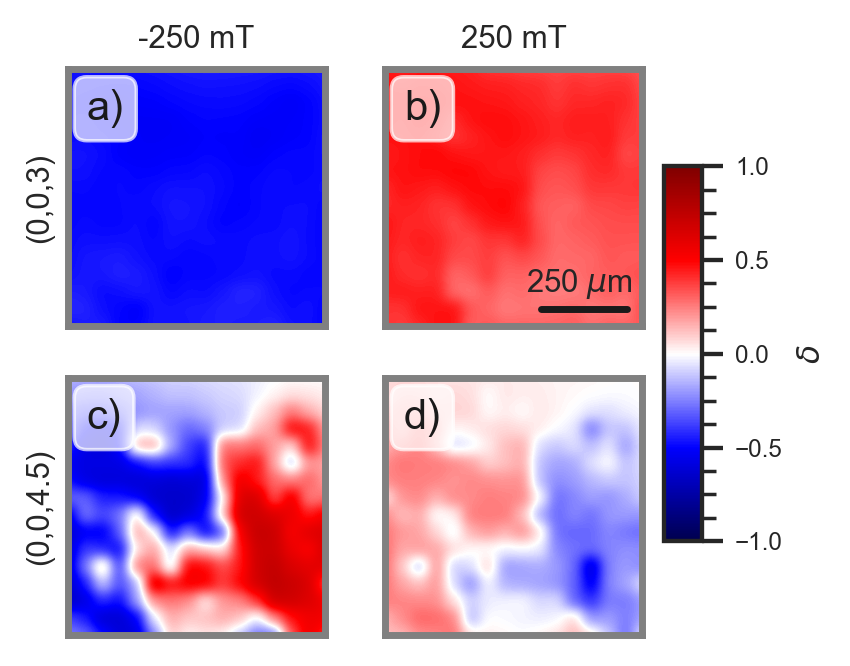}
\caption{Spatially resolved circular dichroism of the (0,0,3) (a and b) and the (0,0,4.5) (c and d) reflections at -250\,mT (a and c) and after isothermal reversal of the magnetic field to +250\,mT (b and d). The contrast of the (0,0,4.5) reflection reflects the magnetic and electric polarity domain configuration of our BSMFO single crystal. The colorbar (far right) is common to all figures and denotes the measured circular dichroism which can take values between $\pm 1$. Figures are to the same scale, denoted in panel b).  In the bottom panels, the left domain (blue in c and red in d) has \emph{positive} toroidal moment, while the right domain (red in c and blue in d) has \emph{negative} toroidal moment (see text).}
\label{fig:CA_domains}
\end{figure}

With a magnetic field of -250\,mT applied to the sample, the (0,0,3) reflection exhibits spatially homogeneous, positive circular dichroism over the extent of the sample (\cref{fig:CA_domains} a), indicating that the sample is uniformly magnetised.

Unlike the (0,0,3) reflection, the spatial map of the (0,0,4.5) reflection exhibits two regions of opposite contrast (\cref{fig:CA_domains} c). Based on \cref{eq: dichroism_calculations_TC}, the two regions of opposite contrast must be magnetic polarity domains with opposite $\lambda$. Importantly, since $\bm{\lambda}$  and $\vec{P}$ are coupled, the two regions must also correspond to \textit{polar} domains of the sample. Because no electrical bias was applied to the sample, we would expect equal populations of the two polarity domains, as we indeed observe.  It is noteworthy that there is a complete lack of correlation between magnetic polarity domains at T = 40 K and B = -250 mT and the zero field helical domains we measured, which are considerably smaller (Figure \ref{fig:helical_domains}). Taking the (0,0,3) and (0,0,4.5) domain maps together, we can conclude that the magnetic polarity domains are also \emph{ferrotoroidal} domains (\cref{eq: ferrotoroidal_definition}).

Next, we investigated the effect of reversing the direction of the applied magnetic field on the domain configuration. Figure \ref{fig:CA_domains}b shows a map of the circular dichroism of the (0,0,3) reflection after field reversal to +250 mT.  The sign of the circular dichroism has switched everywhere on the sample, consistent with a uniform reversal of the sample's magnetisation. Remarkably, the sign of the (0,0,4.5) dichroism has also switched on a pixel-by-pixel basis (\cref{fig:CA_domains} d), i.e., red regions turn blue and blue regions turn red, meaning that the magnetic polarity (and hence the electrical polarisation) has switched sign within each of the domains.  It is noteworthy that the boundary between the two domains has remained pinned throughout the switching process. This observation indicates that the ferrotoroidal domains are very stable, while the magnetisation of each ferrotoroidal domain can be almost freely rotated, resulting in a simultaneous reversal of the electrical polarisation. Finally, we note that the \emph{magnitude} of the (0,0,4.5) dichroism is slightly weaker at +250 mT than at -250 mT (indicated by the fainter colors in the figure). There is therefore a small but noticeable \emph{circular asymmetry} (\cref{eq:circ_asymmetry}), which will be further discussed below.

\begin{figure}
\centering
\includegraphics{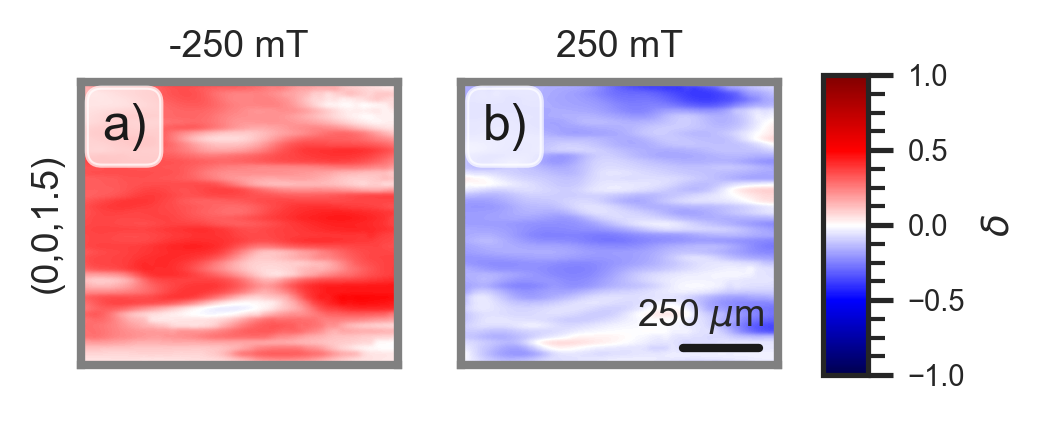}
\caption{Circular dichroism maps of the (0,0,1.5) reflection after an E$_+$B$_+$ magnetoelectric poling procedure. a) for B = -250\,mT and b) for B = +250\,mT.} 
\label{fig:001p5_CA}
\end{figure}

\subsubsection{Ferrotoroidal domain control}

It is well known that ferrotoroidal domains can be manipulated by a combination of orthogonal electric and magnetic fields \cite{baum_2013,spaldin_2008}.  It is therefore natural to assume that electrical biassing of the sample under an applied magnetic field should result in the removal of the ferrotoroidal domain boundaries.  To test this hypothesis, we used a second sample prepared with electrical contacts such that electric fields could be applied perpendicular to the applied magnetic field.  In the absence of electric field bias, the circular asymmetry of the (0,0,1.5) reflection (the (0,0,4.5) reflection was not accessible in this second experiment) exhibited regions of multiple contrast (not shown). However, after performing a magnetoelectric poling procedure identical to the one as described in \cref{section: xtal growth char} (\cref{fig:ME_measurements}), the raster scans of the circular dichroism for the (0,0,1.5) reflection (\cref{fig:001p5_CA} a) became uniform, indicating that the sample had acquired a uniform magnetic polarity.  The magnetisation of the sample is also uniform, as indicated by circular dichroism of the (0,0,3) reflection (not shown here), indicating that the sample now comprises a single ferrotoroidal domain.  Upon reversal of the magnetic field, once again the signs of the circular asymmetry for both (0,0,3) and (0,0,1.5) reflections reverse (\cref{fig:001p5_CA}), indicating a global switching of both magnetisation and magnetic polarity, whilst the direction of the ferrotoroidal moment remains the same --- consistent with our bulk measurements in which the electric polarisation fully switches on inversion of the applied magnetic field (\cref{fig:ME_measurements}).

%We fit our field sweeps, using a step function described by mean, $\alpha$ and difference, $D$. This well describes the `high field' data and we must e an additional field to describe the switching parameters. In terms of the components of the Fourier transform of the spin density for $\sigma$-$\pi$ / $\pi$-$\sigma$ ,

%%\begin{align}
%\alpha  &= |S_3|^2\sin^2(\theta) + |S_1|^2\cos^2(\theta)\\
%D &= \pm 2(S_3S_1^* + S_3^*S_1)\sin(\theta)\cos(\theta)
%\end{align}

%For the $\pi$-$\pi$ channel, in the absence of charge scattering, the difference is necessarily zero and therefore,

%\begin{equation}
%\alpha  = |S_2|^2\sin(2\theta)^2
%\end{equation},
\begin{figure}[!b]
\centering
\includegraphics{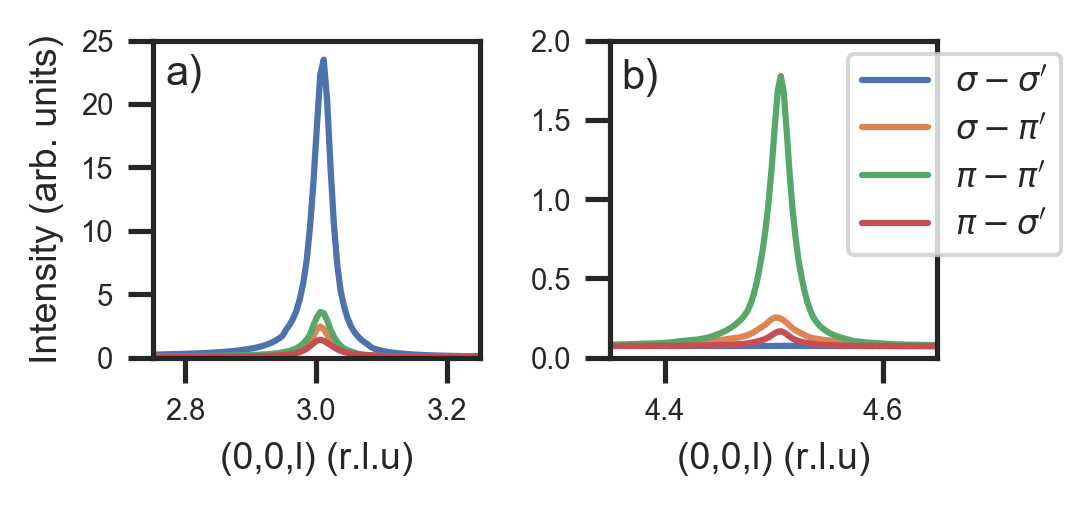}
\caption{Intensity of the (0,0,3) and (0,0,4.5) reflections (a and b respectively) measured in each linear polarisation channel (color).  Measurements were performed at 40\,K and with a 250\,mT magnetic field applied parallel to [2,1,0].}
\label{fig:00l_scans}
\end{figure}

\subsubsection{Linear polarisation analysis}
\label{sec:lin_pol_analysis}

Scans over the (0,0,3) and (0,0,4.5) reflections (TC phase) are displayed in \cref{fig:00l_scans}. Measurements were performed at 40 K and at +250 mT in each of the four linear polarisation channels (defined in the basis of orthogonal linear polarisation states shown in \cref{fig:RXS_geometry}). The (0,0,3) reflection (\cref{fig:00l_scans} a) scatters most intensely in the $\sigma-\sigma'$ channel and is therefore predominately of charge origin, as expected. Weak intensity in both rotated channels ($\sigma-\pi'$ and $\pi-\sigma'$) demonstrates that there is scattering of magnetic origin at this location  (being the only scattering process which can rotate the plane of polarisation of incident light). Conversely, for the (0,0,4.5) reflection (\cref{fig:00l_scans} b) the intensity in the $\sigma-\sigma'$ channel is zero, but scattering is observed in every other linear polarisation channel. This scattering is therefore of \textit{pure} magnetic origin and its properties must, therefore, be linked to the magnetic structure of BSMFO. The importance of this observation becomes apparent when considering that there is a precedent for the commensurate reflections of the Y-type hexaferrites to be of \textit{mixed} charge and magnetic scattering \cite{ueda_2016}.

\subsection{Dynamical measurements during magnetic field switching}

\begin{figure*}
\centering
\includegraphics{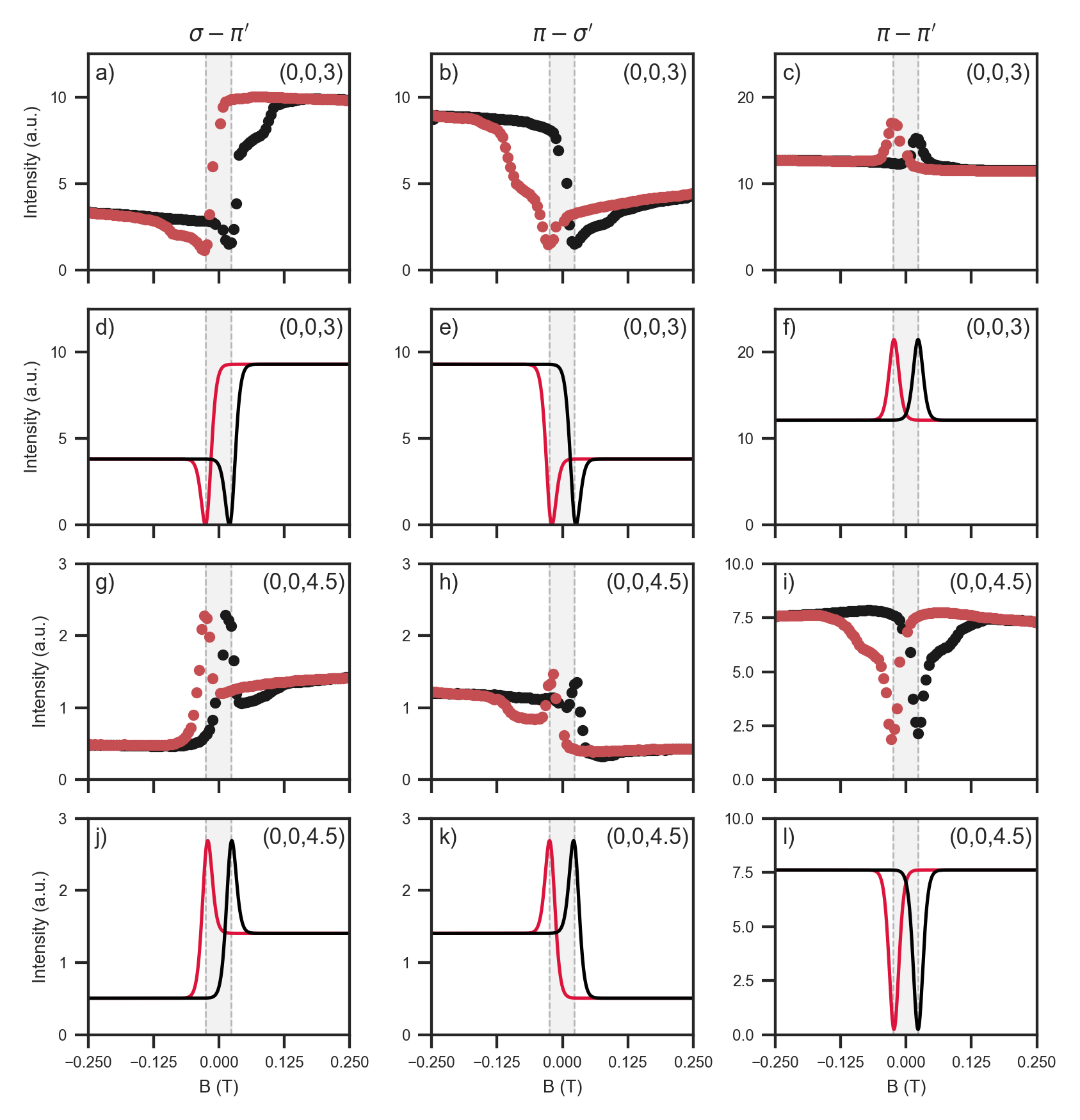}
\caption{a-c) and g-i) Measured scattering intensity for the (0,0,3) and (0,0,4.5) reflections respectively in three linear polarisation channels during the magnetic polarity switching process at 40\,K. The magnetic field was swept at 6.25\,\si{\milli\tesla\per\second}. Black data points show the scattering intensity measured on increasing magnetic field and red points on decreasing. The pair of dashed grey lines are plotted at the same field values (23\,mT and -26\,mT) across all channels. d-f) and j-l) Calculated intensity of the (0,0,3) and (0,0,4.5) reflection respectively, in three linear polarisation channels for the m-TC structure as the magnetic structure is coherently rotated about $c$. Parameters used in the calculations are those discussed in \cref{table: modified TC model output} with the addition of arbitrary scale factors and a parameter controlling the width of the hysteresis during switching. Colour (red or black) denotes the direction of the field sweep (decreasing and increasing field respectively).\newline}
\label{fig:linear_polarisation_switching}
\end{figure*}

The (0,0,3) and (0,0,4.5) diffraction intensities were measured in each linear polarisation channel as the magnetic field was rapidly swept (6.25\,\si{\milli\tesla\per\second}) through several hysteresis loops, enabling one to track how different components of the magnetic structure changed during the switching process.

The most salient features of these data (shown in \cref{fig:linear_polarisation_switching}) are the presence of linear asymmetry (i.e., a step in the scattering intensity between positive and negative magnetic fields),  a large, transient increase or decrease in the scattering intensities occurring at precisely the same field magnitude ($23$\,mT and $-26$\,mT, denoted by dashed grey lines/shaded areas in \cref{fig:linear_polarisation_switching}) for all linear polarisation channels, and significant field hysteresis. Under slow magnetic field switching ($\sim$ 0.1 \si{\milli\tesla\per\second}, not shown) the transient features are greatly suppressed and are therefore to be associated with the dynamic polarity switching process.

Ishiwata \textit{et. al} proposed a mechanism for the magnetic polarity reversal in BMFO (the parent compound of BSMFO), by which the magnetic structure coherently rotates around $\vec{P}$ ($y$), corresponding to a smooth transformation of the TC phase into the LC phase, and then back to a TC phase of opposite polarity in the oppositely applied magnetic field \cite{ishiwata_2008}. Our measurements categorically demonstrate that this cannot be the case in BSMFO. Instead, the peaked intensity in the (0,0,3) $\pi-\pi'$ channel uniquely identifies the evolution of a $y$ component of the magnetisation upon switching, indicating that the magnetic structure rotates about the $c$-axis ($z$) as opposed to $\vec{P}$. Furthermore, a simple calculation of the diffraction intensity in all linear scattering channels based upon the block-model TC structure (not shown), showed that all peaked features observed upon switching are fully consistent with a coherent rotation of the magnetic structure around $c$.

\subsection{\label{sec::beyond_block}Beyond the block model:  the true nature of the TC phase}

Although the peaks near zero field could be accounted for by the block TC model, we emphasise that this model \emph{does not} reproduce the large linear asymmetry (i.e, the `steps' in the scattering intensity) observed in the rotated polarisation channels ($\sigma \pi'$ and $\pi \sigma'$) for both the (0,0,3) and (0,0,4.5) peaks (\cref{fig:linear_polarisation_switching}).  We have carefully considered the possibility that the steps might arise from the $F^{(2)}$ term, which we have thus far ignored (see Appendix \ref{sect_mag_int_rank2} for a detailed calculation of this term).  It is not difficult to show that, for the TC phase, the magnetic interaction rank-2 tensor associated with the (0,0,3) peak  is purely diagonal, and does not contribute to intensity in the rotated channel. A more detailed calculation, reported in Appendix \ref{sec_F2_term_TC} shows that this term does not produce linear asymmetry in the (0,0,4.5) peak either.  Another possibility is that the steps might arise from magnetic field misalignment with respect to the scattering plane.  This possibility can also be ruled out, at least for the simple block-TC model:  the observed asymmetry can only occur if $M_x$ and  $M_z$ are partially in phase, and if only one of them switches upon field reversal (\cref{eq: magnetic_linear_asy}), and this combination can never be obtained with field misalignment for either of the reflections. 
 
Since other possibilities fail, we must come to the conclusion that the observed asymmetry in the rotated channels \emph{must} arise from a sizeable internal distortion of the magnetic structure from the block model considered thus far.  Taking the TC model as a starting point, the simplest way to reproduce the observed steps is to introduce a \emph{scalar modulation} of the magnetic structure with commensurate propagation vector $\vec{k}_Z=$ (0,0,1.5).  In this context, scalar modulation simply means a modulation of the magnetic cone angles and/or of the helical pitch.  Even without performing additional calculations, it is evident that such modulation would \emph{simultaneouly} introduce an $M_z$ component at the $\Gamma$ point (possibly associated with a small net moment in the $z$ direction), \emph{and} an $M_x$ component with $\vec{k}_Z$ propagation vector, both having the correct phase to produce asymmetry in the rotated channels.    We emphasise that the large size of the asymmetry steps means that the $M_x$ and $M_z$ components for both propagation vectors must be of the same order of magnitude, which constitute a significant deviation from the `standard' TC model. 

Indeed, the observed dichroism, asymmetries and switching behaviour are well reproduced (\cref{fig:linear_polarisation_switching} g-i) by such a `modified' 2-fan Transverse Conical model (m-TC), which was constructed with the minimal modifications required to explain the data whilst remaining physical. There are two significant points of departure from the `standard' TC model: 1) The L block is split into two non-collinear sub-blocks, L1 and L2 (representing the bottom and the top of the L block, respectively), making an angle $2 \phi_1$ with each other, and 2) The out-of-plane lifting angle of the S block is no longer constant, but varies by $\pm \psi_1$.  As shown in more detail in Appendix \ref{Appendix: m-TC model}, the effective magnetic moment of each block is constant, and so is the dot product between adjacent blocks, provided that the following constraint is fulfilled:

\begin{equation}
\label{eq: cos_constraint}
\cos \left(\psi -\psi _1\right) \cos \left(\phi +\phi _1\right)=\cos \left(\psi +\psi _1\right) \cos \left(\phi -\phi _1\right)
\end{equation}

Thus, the m-TC model has only the additional parameter $\phi_1$, since $\psi_1$ can be calculated from \cref {eq: cos_constraint}.  The m-TC model has two characteristics making it strongly reminiscent of the zero-field LC phase: it has a net chirality (there are left-handed and right-handed versions of it), and also a small net ferrimagnetic moment along the $z$ axis in addition to the main $x$-axis magnetisation.  In the m-TC model, these two facts are related, since a structure possessing both net chirality and ferrotoroidal moment along $z$ must also, by symmetry, posses a ferromagnetic moment along $z$. 

Table \ref{table: modified TC model output} lists the experimental and calculated values of dichroism and asymmetry in different linear and circular channels for the m-TC model. Since the main fan angle was set to a typical literature value for the TC phase ($\phi=38^{\circ}$) \cite{momozawa_1993}, there are only two variable parameters ($\psi$ and $\phi_1$), which were adjusted to obtain an exact agreement with the (0,0,4.5) experimental values of $\delta_{ave}$ and $Asy_{\sigma \pi'}$.  The agreement we obtain with the other experimental values is quite reasonable, especially considering that the model is over-constrained by the data.   We emphasise again that the deviations from the standard TC model, represented by the angles $\phi_1$ and $\psi_1$, are very large.  Undoubtedly, this is in part a consequence of the simplicity of the model, which fails to capture the full internal complexity of the hexaferrite structure.

\begin{table}
\centering
\caption{\label{table: modified TC model output} Experimental and calculated values of dichroism and asymmetry in different channels for the modified 2-fan transverse conical model (m-TC).  The calculations were performed for a negative-dichroism domain (a `blue' domain in \cref{fig:CA_domains}). The main fan angle was set to $\phi=38^{\circ}$ --- a typical value for the standard TC phase; $\psi$ and $\phi_1$ were adjusted to obtain precise agreement with the (0,0,4.5) experimental values of $\delta_{ave}$ and $Asy_{\sigma \pi'}$, while all the other values are unconstrained outputs of the calculations. Optimal values were  $\psi=36.5^{\circ}$ and $\phi_1=28.0^{\circ}$.  The value of $\psi_1=29.3^{\circ}$ was calculated from \cref{eq: cos_constraint}.  The circular dichroism and asymmetry for the $(0,0,3)$ reflection depend on the unknown ratio $F^{(0)}/F^{(1)}$, and could not therefore be calculated. }
    \begin{ruledtabular}
    \begin{tabular}{c | r  r | r  r }
    &\multicolumn{2}{c|}{Reflection: $(0,0,3)$} & \multicolumn{2}{c}{Reflection: $(0,0,4.5)$}  \\[0.5ex]
    \hline\\[-2ex]
    Parameter&Exp.& Calc. &Exp.& Calc.\\
     \hline\\[-2ex]
     $\delta_{ave}$&$0.49$&--&$-0.59$&$-0.59$\\
     $Asy_{\delta}$&--&--&$0.28$&$0.25$\\
     $Asy_{\sigma \pi'}$&0.52&0.41&$0.47$&$0.47$\\
     $Asy_{\pi \sigma'}$&-0.42&-0.41&$-0.47$&$-0.47$\\
     $Asy_{\pi \pi'}$&-0.04&0&$-0.00$&$0$\
    \end{tabular}
    \end{ruledtabular}
\end{table}

Having only three magnetic Bragg peaks at our disposal (albeit with several polarisation channels), proposing a full model of the `true' magnetic structure of BSMFO in its magnetoelectric phase would be incautious at best. Nevertheless,  we speculate that all non-collinear magnetic structures of Y-type hexaferrites are significantly more complex than is currently believed, and that some of the internal degrees of freedom of these structures are fixed by competing exchanges in a way that is not captured by the block models, and that these features are preserved upon applications of magnetic field, adding unexpected `twists' to the magnetic structure of the magnetoelectric phases.

\section{Conclusions}

We have shown that resonant X-ray microdiffraction can be used to measure the magnetic polarity domain configuration of the Y-type hexaferrite \BSMFO{}. We provide direct evidence of the magnetic polarity switching during inversion of the applied magnetic field and show that electric fields can be used to prepare a single magnetic polarity domain. By dynamic measurement of the scattering during the switching process we were able to propose a mechanism for the polarity switching --- a coherent rotation of the magnetic structure about the $c$ axis, in contention with the switching mechanism originally proposed for the parent compound BMFO \cite{ishiwata_2008}. We also showed that the established magneto-electric structures are inconsistent with our data, and we constructed a minimally modified magnetic structure model that reproduces quantitatively the behaviour of all the X-ray polarisation channels upon field reversal.

%We performed a spatially resolved resonant x-ray diffraction experiment to investigate the polarisation memory effect in the Y-type hexaferrite \BSMFO{}. We observed  charge-magnetic inference at the (0,0,3) bragg reflection, giving us sensitivity to the $\vec{k}$=[0,0,0] component of the magnetic structure, showing that circular asymmetry can be used to image the ferrimagnetic domain structure of a Y-type hexaferrite. We have shown that circular asymmetry on the (0,0,4.5) reflection is of a pure-magnetic origin and can be used as a probe of the magnetic polarity of the system. Using the high-spatial resolution of a national synchrotron source we spatially resolved the magnetic polarity of the system as a function of magnetic field, allowing us to observe it switching on magnetic field reversal. We suggest this reversal occurs by a coherent rotation of the magnetic structure in the $ab$ plane and that the longitiduinal conical state plays no role in the polarity reversal. Our results will enable better interpretation of the Y-type hexaferrite structures in the future.

\section{Acknowledgements}

We acknowledge Diamond Light Source for time on Beam Line I10 under Proposal SI14826 and SI17388. The work done at the University of Oxford is funded by EPSRC Grant No. EP/M020517/1, entitled Oxford Quantum Materials Platform Grant. R. D. J. acknowledges support from a Royal Society University Research Fellowship.  We thank Dr. N. Waterfield Price for discussions and Dr. D. D. Khalyavin for discussions on the m-TC model.

\appendix

\section{Basic aspects of magnetic scattering theory}
\label{Appendix: basic MS theory}

In this section we reproduce the essential, well-known results of the resonant magnetic scattering theory, as it appears, for instance, in J. P. Hill and D. F. McMorrow \cite{hill_1996}, but we recast them in the compact formalism that we employed throughout the paper.

\subsection{X-ray polarization}
Using the cartesian coordinates presented in fig. \ref{fig:RXS_geometry}, we can define the incident and scattered polarisation of the X-rays as follows:

\begin{eqnarray}
\label{eq: x-ray-polarisations}
\epsilon_{\sigma}=\epsilon'_{\sigma}&=&\left( \begin {array}{ccc} 0& 1& 0\end{array}\right)\nonumber\\
\epsilon_{\pi}&=& \left( \begin {array}{ccc}\sin\theta& 0& \cos \theta\end{array}\right)\nonumber\\
\epsilon'_{\pi}&=& \left( \begin {array}{ccc}- \sin \theta& 0& +\cos \theta\end{array}\right)\nonumber\\
\epsilon_{c+}&=&\frac{1}{\sqrt{2} }\left( \begin {array}{ccc}\sin \theta& i& \cos \theta\end{array}\right)\nonumber\\
\epsilon_{c-}&=& \frac{1}{\sqrt{2} }\left( \begin {array}{ccc}\sin \theta& -i& \cos \theta\end{array}\right)
\end{eqnarray}

\subsection{Resonant magnetic x-ray diffraction}
Here, we employ a slightly different formalism with respect to the classic treatment by J. P. Hill and D. F. McMorrow \cite{hill_1996}, with the aim of separating the geometrical terms (which are experiment-dependent) from the magnetic interaction vectors/tensors, which are used extensively in magnetic neutron diffraction dat analysis and can be evaluated once and for all for a given magnetic structure.  For this purpose, we re-write eq. (7) in McMorrow \& Hill \cite{hill_1996} in tensorial form as follows:

\begin{equation}
\label{eq: general_amplitude_expression}
f^{XRES}_{nE1}=\epsilon'_{\alpha}\epsilon_{\beta} T_{tot}^{\alpha \beta}
\end{equation}

where

\begin{equation}
\label{eq: mag_scatt_tensor}
T_{tot}^{\alpha \beta}=F^{(0)} \delta^{\alpha \beta}-iF^{(1)} \epsilon^{\alpha \beta \gamma} M_{\gamma}+F^{(2)}T^{\alpha \beta}
\end{equation}

and the three terms in \cref{eq: mag_scatt_tensor} represent (from left to right) the contributions from Thomson/anomalous charge scattering, $F^{(1)}$ scattering (proportional to the magnetic moments) and $F^{(2)}$ (proportional to the square of the magnetic moments), $F^{(0)}$, $F^{(1)}$ and $F^{(2)}$ being the resonant scattering factors (in general,  complex numbers).  For simplicity, we have included into $F^{(0)}$ the ordinary Thomson structure factor, which is well approximated by a real number (the crystal structure is close to being centrosymmetric).  The values of $\epsilon$ can be taken directly from \cref{eq: x-ray-polarisations} for our specific experimental configurations.  In \cref{eq: mag_scatt_tensor}, $M_{\gamma}$ are the components of the ordinary magnetic interaction vector, a complex vector that also appears in the calculations of magnetic neutron scattering cross sections, so that $\epsilon^{\alpha \beta \gamma} M_{\gamma}$ is an antisymmetric rank-two tensor, while $T^{\alpha \beta}$ is the symmetric magnetic interaction rank-two tensor associated with $F^{(2)}$.  The calculations of  $M_{\gamma}$ and $T^{\alpha \beta}$ for a generic magnetic structure and allowed Bragg peaks are reported in the remainder of this Appendix, for both commensurate and incommensurate magnetic structures.

\subsection{Magnetic interaction vector}

The definition of the complex magnetic interaction vector $\vec{M}$ naturally arises from the calculation of both neutrons and x-rays magnetic scattering cross sections.  For instance, in the case of RXD, the relevant term is 

\begin{equation}
-iF^{(1)} \epsilon^{\alpha \beta \gamma} M_{\gamma}=-i F^{(1) }\left(
\begin{array}{ccc}
 0 & M_z & -M_y \\
 -M_z & 0 & M_x \\
 M_y & -M_x & 0 \\
\end{array}
\right)
\end{equation}

When calculating the scattering amplitude, one needs to perform the following sum, running over the lattice nodes $n$ and the sites in the unit cell $m$:

\begin{equation}
\vec{A}=\sum_{n,m} e^{i \vec{Q} \cdot (\vec{R}_n+\vec{r}_m)}\vec{m}_{nm}
\end{equation}

where the $\vec{m}_{nm}$ are the real magnetic moments at site $m$ in unit cell $n$.  $\vec{R}_n$ denotes the position vector of the origin of unit cell $n$, while $\vec{r}_m$ denotes the position vector of site $m$. Using the propagation vector formalism,

\begin{equation}
\vec{m}_{nm}=\sum_{\vec{k},m}\vec{S}_{\vec{k},m} e^{-i \vec{k} \cdot \vec{R}_n} + \bar{\vec{S}}_{\vec{k},m} e^{+i \vec{k} \cdot \vec{R}_n} 
\end{equation}

where $\vec{S}_{\vec{k},m}$ are the complex-vector Fourier component of the magnetic moment at site $m$ and for propagation vector $\vec{k}$.  After performing the lattice sum, one obtains:

\begin{equation}
\vec{A}=\delta(\vec{Q}-\vec{k}-\vec{\tau})  \vec{M}^+_{\vec{k}}+\delta(\vec{Q}+\vec{k}-\vec{\tau})\vec{M}^-_{\vec{k}}
\end{equation}

where $\vec{Q}$ is the scattering vector, $\bm{\tau}$ is a reciprocal lattice vector and the two magnetic interaction vectors are

\begin{eqnarray}
\vec{M}^+_{\vec{k}}&=&\sum_m e^{i \vec{Q} \cdot \vec{r}_m}\vec{S}_{\vec{k},m}\nonumber\\
\vec{M}^-_{\vec{k}}&=& \sum_m e^{i\vec{Q} \cdot \vec{r}_m} \bar{\vec{S}}_{\vec{k},m}
\end{eqnarray}

In the most general case of $\vec{k}$ within the Brillouin zone, each Bragg peak develops two satellites, one at position $\bm{\tau}+\vec{k}$ and the other at $\bm{\tau}-\vec{k}$.

In the special case of $2 \vec{k}=RLV$ (a reciprocal lattice vector) then both terms will contribute to the same satellite, with the magnetic interaction vector

\begin{equation}
\vec{M}_{\vec{k}}= 2\sum_m e^{i \vec{Q} \cdot \vec{r}_m}\Re(\vec{S}_{\vec{k},m})
\end{equation}

\subsection{Magnetic interaction rank-two tensor}
\label{sect_mag_int_rank2}

Similarly, the scattering amplitude for the $F{(2)}$ term are performed using the magnetic interaction rank-two tensor:

\begin{eqnarray}
B^{\alpha \beta}&=&\delta(\vec{Q}-2\vec{k}-\vec{\tau}) T^{\alpha \beta +}_{2\vec{k}} +\delta(\vec{Q}+2\vec{k}-\vec{\tau}) T^{\alpha \beta -}_{2\vec{k}}\nonumber\\
&&+\delta(\vec{Q}-\vec{\tau}) T^{\alpha \beta}_{0}
\end{eqnarray}

where

\begin{eqnarray}
T^{\alpha \beta+}_{2\vec{k}}&=& \sum_m e^{i \vec{Q} \cdot \vec{r}_m}  \vec{S}_{\vec{k},m}^{\alpha}\vec{S}_{\vec{k},m}^{\beta}\nonumber\\
T^{\alpha \beta-}_{2\vec{k}}&=& \sum_m e^{i \vec{Q} \cdot \vec{r}_m}  \bar{\vec{S}}_{\vec{k},m}^{\alpha}\bar{\vec{S}}_{\vec{k},m}^{\beta}\nonumber\\
T^{\alpha \beta}_{0}&=& \sum_m e^{i \vec{Q} \cdot \vec{r}_m}2 \Re(\vec{S}_{\vec{k},m}^{\alpha}\bar{\vec{S}}_{\vec{k},m}^{\beta} )
\end{eqnarray}

So, there are contributions at $\bm{\tau}\pm 2\vec{k}$ and also at the $\Gamma$ point $\bm{\tau}$.

Once again if $2 \vec{k}=RLV$ then there is only a contribution at the $\Gamma$ point $\bm{\tau}$, with the following magnetic interaction tensor:

\begin{equation}
T^{\alpha \beta}=4\sum_m e^{i \vec{Q} \cdot \vec{r}_m}  \Re(\vec{S}_{\vec{k},m}^{\alpha}) \Re(\vec{S}_{\vec{k},m}^{\beta})
\end{equation}

%( v^{\alpha}_m v^{\beta}_m-w^{\alpha}_m w^{\beta}_m)+i( v^{\alpha}_m w^{\beta}_m+w^{\alpha}_m v^{\beta}_m)
%( v^{\alpha}_m v^{\beta}_m+w^{\alpha}_m w^{\beta}_m)

\section{Magnetic models}
\label{Appendix: magnetic_models}

In this section, we calculate the Fourier components $\vec{S}$, the real magnetic moments $\vec{m}$ and the magnetic interaction vectors $\vec{M}$ for all the phases and reflections of interest for this paper.  We employ either the standard block model with small (S) and large (L) blocks or the modified block model, with the large block split into L1 and L2.  In all cases the values of the magnetic moments depend on the $z$ component of the unit-cell position vector $\vec{R}_n$ (herewith denoted as $R_n$), where $R_n$ can assume all values such that $3 R_n =$ integer.  This is because we employ the hexagonal conventions for the propagation vector, whereas the primitive unit cell is rhombohedral.

\subsection{Helix/LC}

The calculations for the helical and longitudinal-conical (LC) phases are very similar, and are here performed together, the only difference being that the LC phase has a component at the $\Gamma$ point.  The other propagation vector is along the $\Lambda$ line in the Brillouin zone, i.e, $\vec{k}=$(0,0,k$_{hel}$), with k$_{hel}$ in the range 0.6--0.8. The chirality of the helix is denoted by $\gamma$ and is equal to $\pm$ 1.  Herewith, $m^{\perp}_S$ and $m^{\perp}_L$ denote the $z$ axis components of the magnetic moments, while $m^{\parallel}_S$ and $m^{\parallel}_L$ denote the $xy$ plane components.

%Small block
\begin{eqnarray}
\vec{S}^{\Gamma}_S&=&
\left[
\begin{array}{c}
0\\
0\\
 \frac{1}{2} m^{\perp}_S\\
 \end{array}
\right]
\nonumber\\
\vec{S}^{\Lambda}_S&=&
\left[
\begin{array}{c}
 -\frac{1}{2}m^{\parallel}_S \\
 i\gamma\frac{1}{2} m^{\parallel}_S \\
 0\\
\end{array}
\right]
\nonumber\\
\vec{m}_S&=&
\left[
\begin{array}{c}
 -m^{\parallel}_S \cos (2 \pi  k_{hel} R_n) \\
 m^{\parallel}_S \gamma\sin (2 \pi  k_{hel} R_n) \\
 m^{\perp}_S\\
\end{array}
\right]\nonumber\\
\end{eqnarray}

%Large block

\begin{eqnarray}
\vec{S}^{\Gamma}_L&=&
\left[
\begin{array}{c}
 0 \\
 0 \\
 -\frac{1}{2} m^{\perp}_L\\
\end{array}
\right]
\nonumber\\
\vec{S}^{\Lambda}_L&=&
\left[
\begin{array}{c}
 \frac{1}{2} m^{\parallel}_L \left(\cos \left(\frac{\pi  k_{hel}}{3}\right)-i \sin
   \left(\frac{\pi  k_{hel}}{3}\right)\right) \\
 \frac{1}{2} i \gamma m^{\parallel}_L \left(\cos \left(\frac{\pi  k_{hel}}{3}\right)-i \sin
   \left(\frac{\pi  k_{hel}}{3}\right)\right) \\
 0\\
\end{array}
\right]
\nonumber\\
\vec{m}_L&=&
\left[
\begin{array}{c}
 m^{\parallel}_L \cos \left(\frac{1}{3} \pi  k_{hel} (6 R_n+1)\right) \\
 m^{\parallel}_L\gamma\sin \left(\frac{1}{3} \pi  k_{hel} (6 R_n+1)\right) \\
 -m^{\perp}_L \\
\end{array}
\right]
\nonumber\\
\end{eqnarray}

%Magnetic interaction vectors for the three peaks
The magnetic interaction vectors for the three reflections are:
\begin{eqnarray}
\vec{M}_{-k}&=&
\left[
\begin{array}{c}
 -\frac{1}{2} (m^{\parallel}_L+m^{\parallel}_S) \\
 \frac{1}{2} i \gamma(m^{\parallel}_L-m^{\parallel}_S) \\
 0 \\
\end{array}
\right]
\nonumber\\
\vec{M}_{+k}&=&
\left[
\begin{array}{c}
 -\frac{1}{2} (m^{\parallel}_L+m^{\parallel}_S) \\
 -\frac{1}{2} i \gamma(m^{\parallel}_L-m^{\parallel}_S) \\
 0 \\
\end{array}
\right]
\nonumber\\
\vec{M}_{3}&=&
\left[
\begin{array}{c}
 0 \\
 0 \\
 m^{\perp}_S-m^{\perp}_L \\
\end{array}
\right]
\end{eqnarray}

\subsection{TC --- Standard block model}
\label{Appendix: TC model}

The standard TC model has two Fourier components at the $\Gamma$ and $Z$ points of the Brillouin zone, where $\vec{k}_Z=$(0,0,1.5). The main fan angle of the L block is $\phi$, while $\psi$ denotes the out-of-plane lifting angle of the S block.

\begin{eqnarray}
\vec{S}^{\Gamma}_S&=&\left[
\begin{array}{c}
 -\frac{1}{2} m_S \cos \psi  \\
 0 \\
 0 \\
\end{array}
\right]\nonumber\\
\vec{S}^{Z}_S&=&\left[
\begin{array}{c}
 0 \\
 0 \\
 \frac{1}{2} m_S \sin \psi  
\end{array}
\right]\nonumber\\
\vec{m}_S&=&\left[
\begin{array}{c}
 -m_S \cos \psi \\
 0 \\
 m_S \cos (3 \pi  R_n) \sin \psi \\
\end{array}
\right]
\end{eqnarray}

\begin{eqnarray}
\vec{S}^{\Gamma}_L&=&\left[
\begin{array}{c}
 \frac{1}{2} m_L \cos \phi \\
 0 \\
 0 
\end{array}
\right]\nonumber\\
\vec{S}^{Z}_L&&\left[
\begin{array}{c}
 0 \\
 \frac{1}{2} m_L \sin \phi \\
 0 \\
\end{array}
\right]\nonumber\\
\vec{m}_L&=&\left[
\begin{array}{c}
 m_L \cos \phi \\
 m_L \cos (3 \pi  R_n) \sin \phi\\
 0
\end{array}
\right]
\end{eqnarray}

%Magnetic interaction vectors for the three peaks
The magnetic interaction vectors for the three reflections are:

\begin{eqnarray}
\vec{M}_{1.5}&=&
\left[
\begin{array}{c}
 0 \\
 i m_L \sin \phi  \\
m_S \sin \psi  \\
\end{array}
\right]\nonumber\\
\vec{M}_{4.5}&=&
\left[
\begin{array}{c}
 0 \\
 -i m_L \sin \phi  \\
 m_S \sin \psi \\
\end{array}
\right]\nonumber\\
\vec{M}_{3}&=&
\left[
\begin{array}{c}
m_L \cos \phi -m_S \cos \psi \\
 0 \\
 0 \\
\end{array}
\right]
\end{eqnarray}

\subsection{m-TC --- Modified block model}
\label{Appendix: m-TC model}

The modified TC model, described in some detail in the main text, has the same Fourier components of the standard TC model.  The S lifting angle is modulated by $\psi_1$ with modulation wavevector $\vec{k}_Z$.   The L block is split into `lower' L1 and `upper' L2 blocks, each modulated by $\phi_1$ in opposite direction and with modulation wavevector $\vec{k}_Z$.  We obtain,

\begin{eqnarray}
\vec{S}^{\Gamma}_S&=&\left[
\begin{array}{c}
 -\frac{1}{2} m_S \cos \psi  \cos \psi _1 \\
 0 \\
 \frac{1}{2} m_S \cos \psi \sin \psi _1\\
\end{array}
\right]\nonumber\\
\vec{S}^{Z}_S&=&
\left[
\begin{array}{c}
 \frac{1}{2} m_S \sin \psi \sin \psi _1 \\
 0 \\
 \frac{1}{2} m_S \sin \psi  \cos \psi _1 \\
\end{array}
\right]\nonumber\\
\vec{m}_S&=&\left[
\begin{array}{c}
 m_S\left(\cos (3 \pi  R_n) \sin \psi  \sin \psi _1-
   \cos \psi  \cos \psi _1\right) \\
 0 \\
m_S\left(\cos (3 \pi  R_n) \sin \psi \cos \psi _1+\cos \psi \sin
   \psi _1  \right) \\
\end{array}
\right]\nonumber\\
\end{eqnarray}

\begin{eqnarray}
\vec{S}^{\Gamma}_{L1}&=&
\left[
\begin{array}{c}
 \frac{1}{4} m_L \cos \phi  \cos \phi _1\\
-\frac{1}{4} m_L \cos \phi \sin \phi _1   \\
 0 \\
\end{array}
\right]\nonumber\\
\vec{S}^{Z}_{L1}&=&\left[
\begin{array}{c}
 \frac{1}{4} m_L\sin \phi  \sin \phi _1 \\
 \frac{1}{4} m_L \sin \phi  \cos \phi _1 \\
 0 \\
\end{array}
\right]\nonumber\\
\vec{m}_{L1}&=&
\left[
\begin{array}{c}
 \frac{1}{2} m_L \left(\cos (3 \pi  R_n) \sin \phi  \sin \phi
   _1+\cos \phi  \cos \phi _1\right) \\
 \frac{1}{2} m_L \left(\cos (3 \pi  R_n) \sin \phi \cos \phi
   _1-\cos \phi\sin \phi _1  \right) \\
 0 \\
\end{array}
\right]\nonumber\\
\end{eqnarray}

\begin{eqnarray}
\vec{S}^{\Gamma}_{L2}&=&
\left[
\begin{array}{c}
 \frac{1}{4} m_L \cos \phi  \cos \phi _1\\
\frac{1}{4} m_L \cos \phi \sin \phi _1   \\
 0 \\
\end{array}
\right]\nonumber\\
\vec{S}^{Z}_{L2}&=&\left[
\begin{array}{c}
- \frac{1}{4} m_L\sin \phi  \sin \phi _1 \\
 \frac{1}{4} m_L \sin \phi  \cos \phi _1 \\
 0 \\
\end{array}
\right]\nonumber\\
\vec{m}_{L2}&=&
\left[
\begin{array}{c}
 \frac{1}{2} m_L \left(-\cos (3 \pi  R_n) \sin \phi  \sin \phi
   _1+\cos \phi  \cos \phi _1\right) \\
 \frac{1}{2} m_L \left(\cos (3 \pi  R_n) \sin \phi \cos \phi
   _1+\cos \phi\sin \phi _1  \right) \\
 0 \\
\end{array}
\right]\nonumber\\
\end{eqnarray}

From this, by performing explicit dot products, we can easily deduce that the structure has constant moments as a function of $R_n$, and that the angle between L1 and L2 is $2 \phi_1$, and is also unmodulated.  Moreover, in order for the angles between S and L1 in the same unit cell and between  L2 and S in adacent unit cell ($\Delta R_n=1/3$) to be the same, \cref{eq: cos_constraint} must be fulfilled, thereby defining a relation between $\psi_1$ and $\phi_1$.  The magnetic interaction vectors for the three reflections are:

%Magnetic interaction vectors for the three peaks

\begin{eqnarray}
\vec{M}_{1.5}&=&
\left[
\begin{array}{c}
m_S \sin \psi  \sin \psi _1 \\
 i m_L \sin \phi  \cos \phi _1 \\
 m_S \sin \psi \cos \psi _1 \\
\end{array}
\right]
\nonumber\\
\vec{M}_{4.5}&=&
\left[
\begin{array}{c}
m_S \sin \psi  \sin \psi _1 \\
 -i m_L\sin \phi  \cos \phi _1 \\
 m_S \sin \psi  \cos \psi _1 \\
\end{array}
\right]
\nonumber\\
\vec{M}_{3}&=&
\left[
\begin{array}{c}
 m_L \cos \phi  \cos \phi _1-m_S \cos \psi  \cos
   \psi _1 \\
 0 \\
 m_S \cos \psi \sin \psi _1   \\
\end{array}
\right]
\nonumber\\
\end{eqnarray}

\section{$F^{(2)}$ for the TC phase}
\label{sec_F2_term_TC}

In this section, we will demonstrate that the $F^{(2)}$ term cannot produce the observed linear asymmetry for either the (0,0,3) or the (0,0,4.5) peaks.  We need to compute the two magnetic interaction tensors, one for each reflection.  We have:

\begin{eqnarray}
T^{\alpha \beta}_0&=&4\left(
\begin{array}{ccc}
 m_{Lx}^2+m_{Sx}^2& 0 &0 \\
0& m_{Ly}^2 & 0 \\
 0 &0 & m_{Sz}^2 \\
\end{array}
\right)\nonumber\\
T^{\alpha \beta}_k&=&4\left(
\begin{array}{ccc}
 0& \pm i m_{Lx}m_{Ly} &m_{Sx}m_{Sz} \\
\pm i m_{Lx}m_{Ly}& 0 & 0 \\
 m_{Sx}m_{Sz}  &0 &0 \\
\end{array}\right)\nonumber\\
\end{eqnarray}

where $m_{Lx}=m_L \cos \phi$ etc., and the $\pm$ is for the (0,0,1.5) and (0,0,4.5) peaks, respectively

We can immediately see that the term $T^{\alpha \beta}_0$ at the $\Gamma$ point is diagonal, and therefore cannot contribute to scattering in the rotated channels (it is of the same form as an anisotropic Thomson scattering). The term $T^{\alpha \beta}_k$ is completely off-diagonal, and does produce scattering in the rotated channels, but it produces no asymmetry.  To show this, we calculate the scattering cross section for the $\sigma \pi'$ channel as an example:

\begin{eqnarray}
I_{\sigma \pi'}&=&| i \sin \theta (F^{(1)}m_{Sz} \mp F^{(2)} m_{Lx}m_{Ly}|^2\nonumber\\
&=&sin^2 \theta \left( |F^{(1)}|^2 m_{Sz}^2+|F^{(2)}|^2 m_{Lx}^2m_{Ly}^2 \right.\nonumber\\
&& \left. +2  \Re (F^{(1)}\overline{F^{(2)}}m_{Sz}m_{Lx}m_{Ly})\right)
\end{eqnarray}

It is easy to see that this term does not change by rotation of the TC structure by 180$^\circ$ around the $z$ axis.

\end{document}